\documentclass{IEEEtran}

\usepackage[utf8]{inputenc}
\usepackage[T1]{fontenc}
\usepackage{microtype}
\usepackage[nocompress]{cite}

\usepackage{xspace}
\usepackage{relsize} 

\usepackage{amsmath}
\usepackage{amssymb}
\usepackage{stackrel}

\usepackage{paralist}
\usepackage[lowtilde]{url}

\renewcommand{\subsection}[1]{\smallskip\noindent{\bf{#1.}}}
\newcommand{\smallsec}[1]{\smallskip\noindent{\bf\textit{#1.}}}
\renewcommand{\subsubsection}[1]{\smallskip\noindent{\em\bfseries #1.}}

\usepackage{longtable}
\usepackage{booktabs}
\usepackage{multirow}
\usepackage{rotating} 

\usepackage{tabulary}

\usepackage{dcolumn}
\newcolumntype{d}{D{.}{.}{2.1}}
\newcolumntype{e}{D{.}{.}{4.3}}
\newcolumntype{f}{D{.}{.}{3.3}}
\newcolumntype{g}{D{.}{.}{3.1}}
\newcolumntype{h}{D{.}{.}{2.2}}
\newcolumntype{i}{D{.}{.}{4.1}}

\usepackage[table]{xcolor}
\usepackage{graphicx}
\graphicspath{{figures/}}

\usepackage[tight,TABTOPCAP]{subfigure}

\usepackage{tikz}
\usepackage{pgflibraryshapes}
\usetikzlibrary{arrows}
\usetikzlibrary{chains}
\usetikzlibrary{fadings}
\usetikzlibrary{fit}
\usetikzlibrary{positioning}
\usetikzlibrary{shapes}

\usepackage[noend]{algorithmic}
\usepackage{algorithm}

\usepackage{fancyvrb}
\usepackage{listings}
\lstdefinestyle{C}{
    language=C,
    basicstyle=\ttfamily\small,
    numberblanklines=true,
    columns=fixed,
    aboveskip=2pt,
    belowskip=1pt,
    lineskip=0pt,
    numbers=left,
    numberstyle=\tiny,
    numberfirstline=true,
    firstnumber=1,
    xleftmargin=15pt,
    morekeywords={assert},
}

\usepackage[rounded]{syntax}
\let\oldsdlengths\sdlengths
\renewcommand\sdlengths{\oldsdlengths\setlength{\sdfinalskip}{0pt}}

\usepackage[%
  pdfpagemode=UseNone,
  pdfpagelabels=false,plainpages=true,
  ]{hyperref}

\makeatletter
\renewcommand\fs@ruled{\def\@fs@cfont{\bfseries}\let\@fs@capt\floatc@ruled
  \def\@fs@pre{\kern4pt\hrule height.8pt depth0pt \kern2pt}%
  \def\@fs@post{\kern0pt\hrule\relax\vspace{-5mm}}%
  \def\@fs@mid{\kern2pt\hrule\kern2pt}%
  \let\@fs@iftopcapt\iftrue}
\makeatother

\newcommand\definetool[2]{\newcommand{#1}{{\smaller\sc #2}\xspace}}
\definetool{\blast}     {Blast}
\definetool{\cpachecker}{CPAchecker}
\definetool{\cbmc}      {CBMC}
\definetool{\cil}       {Cil}
\definetool{\llvm}      {LLVM}
\definetool{\tvla}      {Tvla}
\definetool{\ocaml}     {OCaml}
\definetool{\tvp}       {Tvp}
\definetool{\camplp}    {CamlP4}
\definetool{\foci}      {Foci}
\definetool{\tcp}       {TCP}
\definetool{\escjava}   {ESC/Java}
\definetool{\slam}      {SLAM}
\definetool{\jpf}       {JPF}
\definetool{\sycmc}     {SyCMC}
\definetool{\impact}    {Impact}
\definetool{\wolverine} {Wolverine}
\definetool{\ufo}       {UFO}

\newcommand{\cegar}{{CEGAR}\xspace}
\newcommand{\safe}{{\sc safe}\xspace}
\newcommand{\unsafe}{{\sc unsafe}\xspace}

\newcommand{\preds}{\mathcal{P}}

\newcommand{\precisions}{\Pi}
\newcommand{\pr}{\pi}

\newcommand{\progpr}{p}

\usepackage{flushend}

\begin{document}

\newcommand{\mytitle}{Reusing Precisions for \\Efficient Regression Verification}
\newcommand{\myauthors}{
Dirk Beyer\,$^1$,
Stefan Löwe\,$^1$,
Evgeny Novikov\,$^2$,
Andreas Stahlbauer\,$^1$,
and Philipp Wendler\,$^{1}$
}
\newcommand{\myaffiliations}{
{$^1$\,University of Passau, Germany}\\
{$^2$\,Institute for System Programming (ISP\,RAS), Russia}
}

\pagestyle{empty}
\hspace{1.5mm}
\begin{minipage}{17cm}
\begin{center}
~\\[3cm]
\Huge{\mytitle}
\\[2cm]
\large{\myauthors}
\\[1cm]
\normalsize
	\myaffiliations\\[5cm]

\vspace{20mm}
\hspace{-5mm}
\includegraphics[scale=0.2]{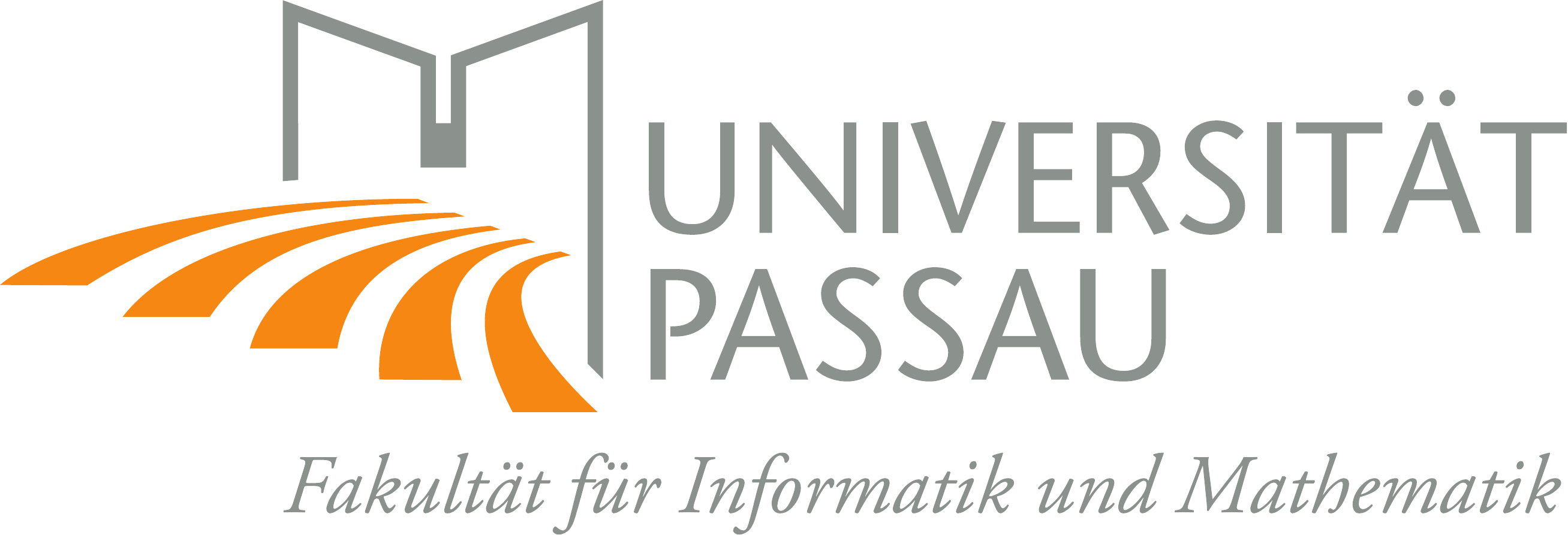} \\[1cm]
Technical Report, Number MIP-1302\\
Department of Computer Science and Mathematics\\
University of Passau, Germany\\
May 2013
\end{center}
\end{minipage}

\title{\mytitle}

\author{
\myauthors
\vspace{1ex}\\
\myaffiliations
}

\begin{titlepage}
\null
\end{titlepage}

\maketitle

\pagenumbering{arabic}
\thispagestyle{empty}
\pagestyle{plain}

\newcommand{\numofdrivers}{59\xspace}
\newcommand{\numofrevisions}{1\,119\xspace} 
\newcommand{\numoftasks}{4\,193\xspace} 
\newcommand{\numofruns}{16\,772\xspace} 
\newcommand{\numofdriverspecs}{259\xspace} 

\begin{abstract}
Continuous testing during development is a well-established technique
for software-quality assurance.
Continuous model checking from revision to revision is not yet established
as a standard practice,
because the enormous resource consumption makes its application impractical.
Model checkers compute a large number of verification facts that
are necessary for verifying if a given specification holds.
We have identified a category of such intermediate results that are
easy to store and efficient to reuse: \emph{abstraction precisions}.
The precision of an abstract domain specifies the level of abstraction
that the analysis works on.
Precisions are thus a precious result of the verification effort and
it is a waste of resources to throw them away after each verification run.
In particular, precisions are small and thus easy to store;
they are easy to process and have a large impact on resource consumption.
We experimentally show the impact of precision reuse 
on industrial verification problems, namely,
\numofdrivers~device drivers
with \numofrevisions~revisions
from the Linux kernel.

\end{abstract}

\section{Introduction}

Reliable software is essential
both for convenience and safety in our daily lives and
for the revenue in the economy.
Producing reliable software is costly;
and speeding up testing and formal verification of software can save huge amounts of time and money.
Economic pressure requires companies to come up with innovations more quickly
by introducing more features in shorter release cycles ---
software is a key contributor to today's innovations.
However, the problem of extending software, e.g., by introducing a new feature,
is that this might break existing features --- bugs get introduced.
This is known as regression.
To avoid regression,
developers execute automated tests
before a new revision of a piece of software is checked-in,
in the hope that the tests alarm the developer of any new bug.
While regression testing is an established and well-investigated technique 
since many years (e.g.,~\cite{Myers:1979,RothermelHarrold:1996,HarroldEtAl:1998}),
in the end, the quality of the software (in terms of correctness) depends
on the coverage percentage achieved by the regression test suite.

The confidence of correctness can be increased by augmenting the development process
with formal verification, i.e., regression verification
\cite{HJMSa03,StrichmanGodlin:2008,HardinEtal:1996,ChakiEtal:2012,SeryEtal:2012}.
Formal verification exhaustively checks the program for bugs,
but at the same time consumes large amounts of computation resources (time and memory),
in particular when applied to industrial-size software.
Regression verification applies formal verification techniques
to continuously check development revisions in order to identify regressions early.
Innovations in this field pave the road
that leads from regression testing to regression verification,
and from simply finding bugs to actual proofs of correctness
during the whole software-development process.

Verification tools spend much effort on computing intermediate results
that are needed for verifying if the specification is satisfied.
In most uses of model checking, these intermediate results are erased
after the verification process
--- wasting precious information (in failing and succeeding runs).
There are several directions to reuse (intermediate) results.
\emph{Conditional model checking}~\cite{ConditionalModelChecking,ChristakisMW12} 
outputs partial verification results
for later re-verification of the same program by another verification approach.
\emph{Regression verification}
\cite{HJMSa03,StrichmanGodlin:2008,HardinEtal:1996,ChakiEtal:2012,SeryEtal:2012}
outputs intermediate results (or checks differences)
in order to enable a more efficient
re-verification of a revised program relying on the very same verification approach.

The contribution of this paper is to reuse \emph{precisions} 
as intermediate verification results.
In program analysis,
e.g., predicate analysis, shape analysis, or interval analysis,
the respective abstract domain defines the kind of abstraction that is
used to automatically construct the abstract model.
The \emph{precision} for an abstract domain defines the level of abstraction
in the abstract model,
for example, which predicates to track in predicate analysis,
or which pointers to track in shape analysis.
Such precisions can be obtained automatically; 
interpolation is an example for a technique that extracts
predicate precisions from infeasible error paths.

Precisions are a good choice for reuse in regression verification, because
they are technically easy to use and do not require much extra computation effort before
they can be reused,
they have a small memory footprint, and they are,
as we show,
not sensitive to changes in the program source code.
We performed an extensive experimental study on industrial code,
in order to show the significant impact
of precision reuse for regression verification (in terms of performance gains).
The benchmark verification tasks were extracted from the Linux kernel,
which is an important application domain~\cite{LDV12},
and prepared for verification using the Linux Driver Verification toolkit~(LDV)~\cite{LDV, LDV-SYSTEM}.
Our study consisted of a total of \numofruns~verification runs for
\numoftasks~verification tasks,
composed from a total of \numofrevisions~revisions
(spanning more than 5~years)
of \numofdrivers~Linux drivers
from the Linux kernel repository.
\begin{table*}[t]
\vspace{-1mm}
\caption{Verification of Linux device driver \texttt{extcon-arizona} without and with precision reuse}
\label{tbl:intro}
\vspace{-1mm}
\centering
\scalebox{0.88}{
\hspace{-2mm}
\begin{tabular}{c@{\hspace{1ex}}rl||l|rr|rr|dr||l|rr|rr|rr}
\toprule
 & Rev.
 & Commit Message 
 & Result 
 & \multicolumn{1}{c}{\begin{sideways} {\bf Refinements} \end{sideways}}
 & \multicolumn{1}{c|}{\begin{sideways} with Reuse \end{sideways}}
 & \multicolumn{1}{c}{\begin{sideways} \!{\bf Abstractions} \end{sideways}}
 & \multicolumn{1}{c|}{\begin{sideways} with Reuse \end{sideways}}
 & \multicolumn{1}{c}{\begin{sideways} {\bf CPU Time} \end{sideways}}
 & \multicolumn{1}{c||}{\begin{sideways} with Reuse \end{sideways}}
 & \multicolumn{1}{c|}{Result}
 & \multicolumn{1}{c}{\begin{sideways} {\bf Refinements} \end{sideways}}
 & \multicolumn{1}{c|}{\begin{sideways} with Reuse \end{sideways}}
 & \multicolumn{1}{c}{\begin{sideways} \!{\bf Abstractions} \end{sideways}}
 & \multicolumn{1}{c|}{\begin{sideways} with Reuse \end{sideways}}
 & \multicolumn{1}{c}{\begin{sideways} {\bf CPU Time} \end{sideways}}
 & \multicolumn{1}{c}{\begin{sideways} with Reuse \end{sideways}}
\\
\midrule
\multirow{10}{*}{\includegraphics[height=3.7cm]{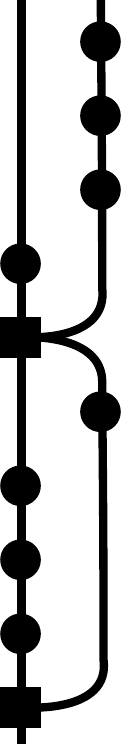}}
&   3 & Implement button detection support   & safe & 24 & 24 & 792 & 792 & 10 & 10 & unsafe & 8  & 8 & 38 & 38 & 3.7 & 3.6 \\ 
&   4 & Free MICDET IRQ on error during probe & safe & 24 & 0  & 792 & 27  & 9.9  & 3.5  & unsafe & 8  & 0 & 38 & 14 & 3.6 & 3.4 \\ 
&   5 & fix typos in extcon-arizona & safe & 24 & 0  & 792 & 27  & 9.8  & 3.5  & unsafe & 8  & 0 & 38 & 14 & 3.6 & 3.4 \\ 
&   6 & Use bypass mode for MICVDD & safe & 4  & 0  & 10  & 3   & 3.2  & 3.1  & unsafe & 1  & 0 & 3 & 2 & 3.1 & 3.1 \\ 
&   7 & Merge tag 'driver-core-3.6' of git://git.kernel.org/\ldots  & safe & 24 & 0  & 792 & 27  & 10 & 3.5  & unsafe & 8  & 0 & 38 & 14 & 3.6 & 3.4 \\ 
&   8 & \textbf{unlock mutex on error path in arizona\_micdet()} & safe & 24 & 0  & 792 & 27  & 10 & 3.5  & safe   & 43 & 16 & 571 & 524 & 8.8 & 8.3 \\ 
&   9 & remove use of \_\_devexit                                   & safe & 24 & 0  & 792 & 27  & 10 & 3.6  & unsafe & 8  & 0 & 38 & 22 & 3.5 & 3.7 \\ 
&  10 & remove use of \_\_devinit                                   & safe & 24 & 0  & 792 & 27  & 10 & 3.5  & unsafe & 8  & 0 & 38 & 22 & 3.5 & 3.8 \\ 
&  11 & remove use of \_\_devexit\_p                                & safe & 24 & 0  & 792 & 27  & 10 & 3.6  & unsafe & 8  & 0 & 38 & 22 & 3.6 & 3.7 \\ 
&  12 & Merge tag 'pull\_req\_20121122' of git://git.kernel.org/\ldots      & safe & 24 & 0  & 792 & 27  & 10 & 3.6  & safe   & 43 & 0 & 571 & 27 & 8.8 & 3.7 \\ 
\midrule

  \multicolumn{3}{c||}{}
& \multicolumn{7}{c||}{Specification 1: `Spinlocks lock/unlock'}
& \multicolumn{7}{c}{Specification 2: `Mutex lock/unlock'} \\
\bottomrule
\end{tabular}
}
\vspace{-4mm}
\end{table*}

\smallsec{Example}
We consider ten revisions of the Linux device driver \verb|extcon-arizona|
for which a bug was discovered using formal verification
by the LDV team%
\footnote{\small\url{https://patchwork.kernel.org/patch/1694901/}}.
Table~\ref{tbl:intro} lists the revisions and the corresponding commit messages
(in bold: the commit that fixes the above mentioned bug).
We verify two specifications with a \cegar-based predicate analysis:
(1)~`Spinlocks lock/unlock',~and (2)~`Mutex lock/unlock'.
Revisions $3$ to $7$ and $9$ to $11$
violate specification~2.
Tasks that violate the specification generally need less refinements
and abstraction computations
since the analysis can terminate as soon as it finds a bug.
In cases where the specification holds,
the whole state space of the program has to be analyzed;
mostly a large number of refinements ($>20$)
and expensive abstraction computations ($>500$) have to be performed.

The columns titled `with Reuse' show the results with precision reuse.
For cases where a complete reusable precision
from a successful verification of a previous revision
is not available
(revision~$3$ for specification~1, revisions $3$ to $8$ for specification~2)
because the whole state space was not yet analyzed before,
there is no speedup.
For most of the cases where the state space was completely analyzed in a previous run,
and a complete precision is available for reuse,
a speedup of at least factor~$2$ can be achieved
(CPU time less than $4$\,s instead of greater than $8$\,s).
Refinements are eliminated completely
because all necessary verification facts
are already specified by the reused precision.

Verifying large numbers of (more complex) program revisions
often takes several hours or even days.
Our approach of precision reuse
can speed this up by a factor greater than~$3$ on average
for predicate analysis.

\subsection{Contributions}
We make the following novel contributions:
\begin{itemize}
\item We identify the abstraction precisions as intermediate results that are valuable
       for reuse in regression checking.
\item We define a tool-independent format for persistent storage and exchange of precisions.
\item We extend an existing software-verification tool in order to support
       regression verification with precision reuse.
\item We prepare and consolidate a benchmark set for regression verification
       that is based on industrial source code from the Linux kernel and consists of thousands 
       of benchmarks.
\item In an extensive experimental study, we show that precision reuse leads to significant
       performance improvements and causes almost no overhead for the verification tool as
       well as for the benchmarking infrastructure (and thus, forms no additional barriers
       in a software-development process).
\end{itemize}

\subsection{Related Work}
The goal of constructing efficient tools for incremental formal verification
is more than 15 years old~\cite{SokolskyScott:1994, HardinEtal:1996}.
In the literature, there exist two main directions to approach
the problem of regression verification:
(1) based on analyzing the difference
between the program and other programs
that were successfully verified in a previous verification run,
and (2) based on reuse of intermediate results
that were costly computed in previous verification runs.

\subsubsection{Verification of Differences}
The first group of approaches to efficient regression verification
takes two programs as input and analyzes the differences
in order to verify whether the specification is still fulfilled.
An input condition is used to limit the verification
to certain relevant parts of the state space~\cite{GodlinStrichman:2009, BoehmeEtal:2013}.
These approaches can be seen as conditional model checking~\cite{ConditionalModelChecking},
where the input condition instructs the verifier to perform a partial verification.
The parts of the program that were identified
as not being affected by modifications can be skipped~\cite{GodlinStrichman:2009, PersonEtal:2011}
during the verification process. 
A technique
for proving conditional equivalence of two programs~\cite{GodlinStrichman:2009}
isolates and abstracts the functions of both versions
using uninterpreted functions and then proves their equivalence
(also extended to multi-threaded programs~\cite{ChakiEtal:2012}).

\subsubsection{Reuse of Verification Results}
The other group of approaches reuses state-space graphs~\cite{LauterburgEtal:2008, HJMSa03},
constraint solving results~\cite{YangEtal:2012, VisserEtal:2012},
or function summaries~\cite{SeryEtal:2012}.
To ensure that the information is valid to be reused,
those parts of the information that were affected by modifications
(to the analyzed program or its specification)
have to be validated.
The check for reusability is done either
before the actual formal verification process is started%
~\cite{SeryEtal:2012, YangDR09, YangEtal:2012}
or immediately before certain information should be reused%
~\cite{HJMSa03, LauterburgEtal:2008}.
Extreme model checking~\cite{HJMSa03} is the only existing approach
that uses un-bounded model checking
with lazy abstraction and predicate analysis
for regression verification.
Another way of information reuse is to not store the concrete data, but its hash value.
One such approach~\cite{HardinEtal:1996} stores hashes of verified models;
these models are constructed by reducing a program
to those parts that are relevant to prove one property.
To be efficient, model construction must be less expensive than verifying the model.
For formal regression verification of hardware using the ic3 algorithm,
the reuse of correctness proves and counterexamples has been proposed~\cite{ChocklerEtal:2011}.
A more general fashion of reuse is
to store and reuse canonicalized constraint solver queries
and the corresponding results.
This idea is supported by the Green framework~\cite{VisserEtal:2012},
which provides a solver wrapper interface.

Our approach belongs to this second category:
we reuse abstraction precisions as intermediate results
and do not (explicitly) analyze the differences in the program code
(our approach implicitly spends more effort on changed parts).
This is the first work that reuses abstraction precisions.

\section{Background}
\subsection{Abstract Reachability Graph}
The class of analyses we consider in our work
is based on creating an abstract model of the program
in form of an abstract reachability graph~(ARG).
An example for such an analysis is implemented in \blast~\cite{BLAST}.
The ARG is created iteratively
by unrolling the control-flow automaton~(CFA) of the program,
creating an abstract successor state for the next location
whenever the control flow passes through an edge of the CFA.
The creation of abstract-successor states is usually over-approximating
and guided by some form of precision that instructs the analysis
which facts should be tracked and which facts should be omitted by abstraction.
The abstract domain determines the characteristics of the precision.
For example, if the abstract domain tracks information of program variables explicitly,
then the set of relevant program variables to consider at a program location
is a suitable precision for the analysis.
The precision in use should require the tracking of just enough information to prevent false alarms,
while at the same time be as concise as possible in order to enable an efficient analysis.

\subsection{Counterexample-Guided Abstraction Refinement (\cegar)}
\cegar~\cite{ClarkeCEGAR} is a well established technique
for automatically finding a suitable precision that matches the above criteria.
Beginning with an initial coarse or even empty precision,
the ARG is created based on this initial precision.
If no state violating the specification is found,
the program is proved safe.
If a violation of the specification is found,
the concrete path of this counterexample is analyzed for feasibility.
If it is feasible, the program is unsafe and the analysis terminates.
Otherwise
the abstract model of the program was too coarse,
so the precision needs to be refined
to exclude this infeasible counterexample from future explorations.
Depending on the abstract domain,
the facts necessary to rule out this counterexample
are extracted from the proof of infeasibility and added to the precision.
Then the \cegar loop is restarted with this newly refined precision.

\subsection{Lazy Abstraction}
The efficiency of \cegar-based analyses
can be increased by using
lazy abstraction~\cite{LazyAbstraction}.
Instead of always restarting the analysis from scratch
after an infeasible counterexample was found,
the abstract model is refined in a ``lazy'' style.
That is, during counterexample analysis the newly-learned facts that are
extracted from the counterexample are only added where necessary.
Then only those parts from the ARG 
that were computed with a too coarse precision
are removed and scheduled for re-exploration.
The remainder of the ARG,
for example, a prefix of the current counterexample path,
or other paths not related to the current counterexample,
are kept and are neither thrown away nor re-explored.
This does not only reduce unnecessary recomputations,
but also reduces computation effort by lazily applying the new, stronger, precision
only to those states of the ARG where it is needed.
States on unrelated paths of the ARG are still computed
with the old, weaker, and more efficient, precision.
A further improvement is to use different precisions
for each program location in order to track as little information as possible.
For example, the analysis drops information during path exploration
when reaching a location after which this piece of information is no longer needed.

\subsection{Predicate Analysis}
One technique which is used widely together with the above concepts
is predicate abstraction~\cite{GrafSaidi97}.
Given the set~$X$ of program variables,
and the set~$\preds$ of quantifier-free predicates over variables from~$X$,
the abstract domain here is the set of
boolean combinations of predicates from~$\preds$.
The precision~$\pi$ is a set of predicates from~$\preds$.
When constructing the ARG,
abstract successor states are created by computing
either the cartesian or the boolean abstraction of the current state
using the predicates from~$\pi$ with an SMT solver.
Using Craig interpolation,
predicates can be generated fully automatically
from a proof of unsatisfiability
for the formula representing a concrete counterexample~\cite{AbstractionsFromProofs}.

The performance of predicate abstraction can be improved
with adjustable-block encoding~(ABE)~\cite{ABE}.
This technique groups program statements into blocks
and computes abstractions only at the end of each block
instead of at all program locations.
Furthermore, if control flow merges within a block,
paths in the ARG are also merged so that sets of paths are considered
instead of single program paths.
When using ABE-Loops
(which encodes loop-free parts of the program into blocks),
abstractions will be computed only at loop-head locations.
Thus predicates will be relevant only at these locations,
and the precision is ignored at all other locations.

\subsection{Explicit-Value Analysis}
Another domain that can utilize a precision is
explicit-state analysis \cite{CPAexplicit},
which tracks the current value for each program variable explicitly.
Within this analysis,
an abstract state is represented as an abstract variable 
assignment~$X \to \mathbb{Z} \mathrel{\cup} \{\top, \bot\}$,
where $X$ denotes the set of program variables of a program.
The value $\top$ represents a variable valuation that is unknown,
e.g., due to an uninitialized variable;
the value $\bot$ represents a variable valuation that is impossible.
Abstract successor computation is done by evaluating program operations
and assigning the evaluated value to the respective program variables
in abstract variable assignments explicitly
--- in contrast to modeling them symbolically as done in the predicate domain.

The precision for an abstract variable assignment
is defined as a set $\pi$ of variables,
which is used to restrict an abstract variable assignment
to variables that are in that precision $\pi$.
For example, applying the precision $\pi = \{b\}$
on the abstract variable assignment $v = \{a \mapsto 4, b \mapsto 15\}$
would result in the abstract variable assignment $v^\pi = \{b \mapsto 15\}$.
Experiments show that a variable that is relevant for one path,
is often relevant on similar paths as well,
and thus it is beneficial to add a newly-found relevant variable
to the precision for all locations of the functions in which it is relevant.
This reduces the number of refinements,
because similar paths can now often also be ruled out
without further refinements.

\section{Precision Reuse}

\subsection{Definitions}
A \emph{precision} is the information
an abstraction-based analysis uses to guide the abstraction
computation for creating abstract states.
Given one analysis, we write~$\precisions$ for the set of possible precisions,
and $\pr$ for one element thereof.
The \emph{empty precision} is the coarsest precision from~$\precisions$
(usually, this precision defines that all information is abstracted).
The \emph{union} of two precisions
from~$\precisions$ is defined in the intuitive way.
For example, for predicate abstraction,
a precision is a set of predicates over program variables,
and the union of two precisions is the union of the two sets of predicates.

In order to use lazy abstraction, which enables the use
of different precisions at different program locations,
we define a \emph{program precision} as a mapping $L \to \precisions$
from the set~$L$ of program locations to the set of precisions~$\precisions$.
The union of two program precisions $\progpr_1$ and $\progpr_2$
is the program precision that maps every location~$l$ to the 
union of $\progpr_1(l)$ and $\progpr_2(l)$.

\begin{figure}[t]
\small
\begin{grammar}
<program-precision> ::= \[[ <header> `\\n\\n' \\
\begin{rep}[b]
    \begin{rep} <scope-selector> \\ ` ' \end{rep} `:\\n' <precision> \\
    `\\n\\n'
  \end{rep}
\]]
<scope-selector> ::= \[[ \begin{stack} `*' \\ "function name" \\ "location number" \end{stack} \]]
<explicit-header> ::= \[[ \]]
<explicit-precision> ::= \[[ \begin{rep}[b] "program variable" \\ `\\n' \end{rep} \]]
<predicate-header> ::= \[[ \begin{rep}[b] <SMT-LIB 2 declaration> \\ `\\n' \end{rep} \]]
<predicate-precision> ::= \[[ \begin{rep}[b] <SMT-LIB 2 assert> \\ `\\n' \end{rep} \]]
\end{grammar}
\vspace{3mm}
\caption{Format of program-precision file; depending on the abstract domain,
          we use the respective definitions of \synt{header} and \synt{precision}}
\label{fig:prec-file}
\vspace{5mm}
\end{figure}

\subsection{Format for Program-Precision Files}
\label{sec:file-format}
In order to write and read precisions to and from persistent storage,
we define a simple text-based file format
that describes program precisions in a human-readable and tool-independent way.
A formal definition of the format is given in the syntax diagram of Fig.~\ref{fig:prec-file}.
The basic structure defined by \synt{program-precision} is the same for all analyses.
The file starts with a header
(the content depends on the analysis)
followed by a blank line.
After the header, an arbitrary number of blocks separated by blank lines follow,
each consisting of one line with a non-empty sequence of scope selectors,
and an analysis-dependent precision.
There are three kinds of scope selectors:
the literal~\verb|*| (representing all program locations),
the name of a function of the program (representing all locations inside this function),
and the number of a program location (representing this single location).
The precision given in a block
is used at all locations represented by the specified scope selectors.
The effective precision for any given program location
is the union over the precisions from all blocks
that contain at least one scope selector representing
that location.

For explicit-value analysis,
the header \synt{explicit-header} is empty.
The precision \synt{explicit-precision}
is a list of variables that occur in the program.
For predicate analysis,
we define the precision \synt{predicate-precision}
as a list of \verb|assert| commands
as defined by the SMT-LIB 2 standard~\cite{SMTLIB2}
(a standard for SMT-solver interfaces
supported by state-of-the-art SMT solvers).
The header \synt{predicate-header} is a sequence of
term declaration commands as defined by SMT-LIB 2
(including the commands \verb|define-fun| and \verb|declare-fun|),
and the formulas in the precisions
may reference these declarations.

\begin{figure}[t]
\small
\hfill
\begin{minipage}[t]{1.5cm}
\begin{Verbatim}[frame=single,fontsize=\footnotesize]
*:
lock

main f:
x
\end{Verbatim}
\end{minipage}
\hfill
\begin{minipage}[t]{6.0cm}
\begin{Verbatim}[frame=single,fontsize=\footnotesize]
(declare-fun |lock|() Real)
(declare-fun |x|() Real)
(define-fun t1() Bool (= |lock| 0))
(define-fun t2() Bool (<= |x| 1))

*:
(assert t1)

main f:
(assert t2)
\end{Verbatim}
\end{minipage}
\hfill
\vspace{5mm}
\caption{Example program-precision files;
left: explicit-value analysis; right: predicate analysis}
\label{fig:example-prec}
\vspace{5mm}
\end{figure}

\smallsec{Example}
Consider a C program that contains two variables \verb|lock| and \verb|x|,
both of which are relevant for proving the safety of the program.
Variable \verb|lock| may be relevant at all locations,
whereas variable \verb|x| may be relevant
only in the functions \verb|main| and \verb|f|.
An example program-precision file for explicit-value analysis
that encodes this information is given on the left in Fig.~\ref{fig:example-prec}.
An example for predicate analysis could look as shown on the right in Fig.~\ref{fig:example-prec},
assuming that the model checker encodes these variables as real numbers,
and the predicates $lock = 0$ and $x \leq 1$ are relevant.

\subsection{Generating Program-Precision Files}
In order to enable the reuse of precisions,
we collect all program precisions that are created during the analysis
(typically, there is one for each refinement step),
and create the union over all these program precisions.
The resulting precision is written to a file.
For each program location,
the precision is dumped and labeled
with the number and the function name of the program location
(empty precisions can be omitted.)

\subsection{Reuse of Precisions}
In order to reuse a precision for the subsequent analysis of the same or a similar program,
an initial program precision for the analysis is created
by interpreting the contents of a previously stored program-precision file.
There are three possibilities to construct such a precision.
First, precisions can be function-scoped,
such that a precision is created for each function of the program,
by taking the union of all precisions labeled with the function name.
The result is assigned to all locations
of the respective function.
Note that this will widen the scope of precisions
(thus potentially leading to a more precise abstraction),
and also loose precisions if functions are renamed.
This precision assignment is insensitive to changes of 
the control-flow structure in the functions of a program.
Second, precisions can be location-scoped
such that the location numbers in the file are read
as the keys for the resulting program precision.
For all program locations that do not appear in the file,
the empty precision is used.
Note that location numbers may change if the program code changes,
and thus, precisions get assigned to locations
that correspond to a semantically different location of the original program.
Third, precisions can be global-scoped
by taking the union of all precisions in the file
and assigning the result to all locations of the program.
This will not loose any precision from the previous analysis,
but might apply precisions to locations where they are not necessary
(and thus make the analysis more expensive).

After the creation of the initial program precision,
the analysis is started as usual.
No changes to the analysis itself are necessary.
If the provided precision is strong enough to prove the program safe,
no further refinement effort will be needed.
If the input precision contains only
a part of the necessary precision to be tracked,
spurious counterexamples will be detected and subsequent refinements
will strengthen the precision.
Note that even in this case the input precision likely
reduces the effort
by decreasing the number of necessary refinements.
This process may be iterated by writing again
the program precision that was further refined
by the second analysis to file,
and using this as the input for a further analysis,
possibly on a newer version of the program.

\subsection{Discussion}
The most significant effect of reusing precisions
from a previous verification run
is the reduction in the number of necessary refinements.
These are usually among the most expensive operations
executed by a model checker
(for example involving satisfiability checks
and interpolation queries over
formulas that represent sets of complete program paths
from the entry point to the error state).
Furthermore,
fewer refinements reduce the number of operations
to prune and re-create parts of the abstract reachability graph.
This is especially important for analyses
that perform expensive operations during this phase,
for example for predicate analysis,
which needs SMT-solver queries to compute abstractions.
While the introduction of adjustable-block encoding~\cite{ABE}
has reduced the number of such computations
by executing them only at loop-head locations
and not for every abstract state,
the need to use boolean abstraction still makes this costly.

Precision reuse is an elegant
and conceptually simple approach,
because it integrates naturally into the techniques
that are used by many successful model checkers.
These techniques can be applied as they exist without any change,
to the first, initial, verification run
(when no reusable information is present),
and also to the subsequent re-verification runs.
Furthermore, this makes precision reuse
applicable not only for the two presented analyses,
but also for any analysis and abstract domain
that is based on \cegar and incorporates an abstraction step
that is guided by some form of precision.
For example, precision reuse would extend naturally
to other abstract domains such as interval or shape analysis.

Precision reuse is easy to implement
in existing model checkers that are based on \cegar and abstractions.
Only the import and export of precisions
before and after the actual analysis
needs to be added.
Complex algorithms,
as required for comparing two revisions of a program
and detecting similar and changed code,
are not necessary in our approach.
The format we defined is easy to parse and write,
and could be supported by a variety of model checkers,
thus even enabling the reuse of precisions
across different tools.

Furthermore, precision reuse is also user-friendly:
a user that is already familiar with using one model checker
will not need to learn how to use new concepts or tools.
Dumping of precisions as part of the analysis result
should be enabled by default in most tools,
and thus the only necessary action by the user
is to supply the previously written program-precision file
as an additional input to the next verification run.
Even if the user mistakenly specifies a wrong program-precision file as input,
the result will be still correct
(the analysis is still sound)
and only the performance might be slightly worse.
In order to employ precision reuse,
it is not necessary to have access to previous program revisions;
the only information needed is the (small) generated program-precision file.

\subsection{Applicability of Precisions}
As described above,
there are three strategies how the precisions from the previous verification run
can be applied to the program locations of the program's next revision.
The strategies differ in how they widen the scope of the precisions.
A location-scoped precision is applied
at exactly those locations
stated in this precision,
risking to not have a precision
at a location where it would be relevant in the new revision.
For example, consider a precision that is relevant
for locations $5$~to~$10$ of a program.
Now, a change is made to the program,
and a statement that is unrelated to the safety of the program
is introduced right after location~$6$.
Thus, the previous locations $5$~to~$10$
now correspond to the new locations $5,6,8-11$.
The previous precision is not applied to location~$11$
and the analysis first fails to prove the program safe,
thus needing at least one additional refinement
to rediscover the missing facts.
Function-scoped precisions
are insensitive to such changes.
Even changes due to cross-cutting concerns
that affect code locally in many functions
are expected to be verifiable without further refinements.
Changes to the call graph of the program,
however, might still generate a similar need for refinements,
for example, if code that is relevant to the safety
of a program is moved to another function.
Global-scoped precisions reduce this problem further,
making refinements only necessary
if code referenced by the precision is changed directly 
(for example, if variables are renamed).

We consider location-scoped precisions to be too sensitive when program code changes.
Which of the other two strategies performs better
depends on the class of program changes
(e.g., whether heavy refactorings changing the functions of the program are common),
and how expensive an unnecessarily coarse precision is for the analysis.
Often, the latter has less effect than one would intuitively consider.
For example, specifying variables from a function~$f$
in the precision of a function~$g$
would have no effect as the variables in $f$ are out of scope in $g$ anyway.
The policy of most projects
is to create small commits with mostly local changes,
thus, we expect function-scoped precisions
to be most promising in practice.

\section{Experimental Evaluation}
In order to evaluate the impact of precision reuse on the effectiveness
and efficiency of regression verification,
we performed an extensive experimental evaluation.
We use industrial software for our experiments:
in total, we prepared \numoftasks verification tasks
from \numofrevisions~revisions of \numofdrivers~device drivers
from the Linux kernel.
We started verification runs on all those problems with both an explicit-value analysis 
and a predicate analysis, each with and without precision reuse.
Our tool implementation, the C source code of the device drivers,
and the full benchmark results
are available on our supplementary web page:
{\small\url{http://www.sosy-lab.org/~dbeyer/cpa-reuse/}}.
During our experiments,
we found an actual bug in the Linux kernel%
\footnote{\small\url{https://lkml.org/lkml/2013/3/1/550}}%
.

\subsection{Implementation}
Our implementation is based on the open-source verification framework \cpachecker{}%
\footnote{\small\url{http://cpachecker.sosy-lab.org}}%
~\cite{CPACHECKER},
which is available under the Apache~2.0 license.
\cpachecker provides implementations
of explicit-state analysis~\cite{CPAexplicit}
and predicate analysis with ABE~\cite{ABE}.
Both approaches are based on \cegar and use a precision to define the level of abstraction.
Thus we only had to add support for writing the program precision to file after a verification run,
and reading in a previously written program precision
to be used as initial precision
before a verification run.
The format for persistent storage of the program precision is described in Sect.~\ref{sec:file-format}.
Further changes to the verification tool were not necessary,
in particular, the verification algorithm and the abstract domains were not changed.
Our extension for precision reuse is integrated into the trunk
of the project's source-code repository%
\footnote{\small\url{https://svn.sosy-lab.org/software/cpachecker}}.

\subsection{Verification Tasks}
A verification task is a fully specified verification input,
which is referred to by a triple that consists of the
name of the driver, the specification that the driver has to satisfy,
and the revision number from the repository.

\subsubsection{Preparation of an Industrial Benchmark for Regression Verification}
We started our selection process by considering the verification tasks
from the category `DeviceDrivers64'
of the 2nd Intl.\ Competition on Software Verification (SV-COMP'13)%
~\cite{SVCOMP13},
which is a benchmark set that consists of 1\,237~verification tasks.
From this set of verification tasks, we selected those device drivers
that fulfill the following two criteria:
(1) \cpachecker, in revision $7481$,
    needed more than 20\,s of CPU time to report either \safe or \unsafe
    (to ensure that the startup time like JVM startup, parsing, etc.,
     does not influence the total run-time too much);
(2) the device driver needs at least one refinement during verification
    (to omit trivial problems and those for which precisions are not needed).

This selection process resulted in a total of \numofdrivers device drivers 
from the SV-COMP'13 benchmarks that fulfilled the above criteria.
We extracted the sources for all available revisions
of those drivers
from the official Linux kernel repository%
\footnote{\scriptsize\url{git://git.kernel.org/pub/scm/linux/kernel/git/torvalds/linux.git}\!\!\!\!}.
Each of these device drivers consists of several header and source files,
each having its own revision history.
We considered all commits
to all C~source files of the device driver,
in chronological order,
starting with the revision in which the device driver was added to its 
directory in the kernel repository
(if the driver resided in the ``staging'' area
of the kernel before being accepted into the main area,
these revisions were not considered).
In order to obtain a linear history of changes
we excluded commits that occurred on
branches that were created during the development of a driver
(the merge commits that reintegrated such branches
 are included, and thus no changes are lost).
The oldest revisions taken date back to the year~2007,
and the latest ones to the end of~2012.

\begin{table}
\caption{Considered specifications (LDV rules)\protect\footnotemark}
\centering
\vspace{1mm}
\small
\begin{tabulary}{\linewidth}{@{}l@{\hspace{1.9ex}}J@{}}
Name & Description \\
\toprule
08\_1a  &\mbox{\emph{Module get/put.}}
  For each successful call to try\_module\_get()
  a corresponding call to module\_put() that unblocks the module must exist. \\
32\_1   &\mbox{\emph{Mutex lock/unlock.}}
  A less accurate implementation of specification 32\_7a. \\
32\_7a  &\mbox{\emph{Mutex lock/unlock.}}
  A mutex must not be acquired or released twice.
  A mutex must not be released without prior acquiring.
  Finally, all mutexes must be released. \\
39\_7a  &\mbox{\emph{Spinlocks lock/unlock.}}
  A spin lock must not be acquired or released twice.
  A spin lock must not be released without prior acquiring.
  Finally, all spin locks must be released. \\
43\_1a  &\mbox{\emph{Memory allocation inside spinlocks.}}
  The flag for atomic allocation operations must be used
  whenever a memory allocation function call is done while a spin lock is held. \\
68\_1   &\mbox{\emph{USB alloc/free urb.}}
  For each allocation of an USB Request Block (URB) using usb\_alloc\_urb()
  a corresponding call to usb\_free\_urb() must exist. \\
\bottomrule
\label{table:ldv-rules}
\end{tabulary}
\end{table}
\footnotetext{\small\url{http://linuxtesting.org/ldv/online?action=rules}}

In order to obtain verification tasks, we also need specifications.
We used as specification six different rules for correct Linux kernel core API usage
(see Table \ref{table:ldv-rules}).
We composed each revision of the \numofdrivers~selected drivers 
with each specification.
The composition was done using the LDV-toolkit\,%
\footnote{\small\url{http://linuxtesting.org/project/ldv}}~\cite{LDV-SYSTEM,LDV}
and consisted of:
(1)~adding a main function that simulates calls to the device driver
from the Linux kernel core,
(2)~weaving in one of the six specifications (reducing the rule-based
specification of the property into a reachability property by weaving in a monitor automaton),
and
(3)~combining all files that the device driver (in the particular revision)
consists of, into a single file (using \cil pre-processing).
The result of this composition process is a verification task
that consists of a single verifiable C file,
for each revision.

We omitted tasks where the specification is trivially satisfied,
e.g., specification ``Module get/put''
for drivers
that do not call the function try\_module\_get().
For evaluating the effect of our approach, we need to consider those verification
tasks for which the precision needs to be fully discovered and where
repeated application of the verifier yields deterministically the same precision.
This is not the case for verification tasks with a known specification violation,
because the analysis can terminate as soon as finding a counterexample,
skipping parts of the state space.
Of course, precision reuse is applicable in such cases as well
(witnessed by the bug we found),
but in our benchmarks the numbers would not be comparable.
Therefore, we remove from our benchmark set all verification tasks with
the expected result \unsafe.
The resulting benchmark set for regression verification
consists of a total of \numoftasks~industrial-strength verification tasks,
which allows us to perform a significant experimental study.

\subsubsection{Differences between Verification-Task Revisions}
While normally source-code changes
for the device drivers are rather limited
from revision to revision,
our benchmark set has quite large source-code differences
between revisions,
which is (not by design, rather as a side-effect) good to evaluate insensitivity to changes.
We explain the main three reasons in the following:
(1)~Whenever commits occurred in branches,
we did not include the corresponding revisions along the branch;
instead, we extracted only revisions from the mainline branch.
The revisions after a merge into the mainline branch
result from a single (generally larger) commit.
(2)~Another reason for a large difference between revisions is
the omission of revisions with a known specification violation.
Thus, the changes from such revisions appear together with the changes
of the next commit, in the succeeding revision without a specification violation.
(3)~Another cause for large differences is that we took one snapshot of the code
for each revision in which one of the actual core device-driver source files changed.
However, in the kernel project there are many other (header) files that influence
the code of a particular file, by being included from the file,
and by defining macros, types, inline functions etc., which are used in the code.
Thus the change between two revisions incorporates not only the changes
of the actual device-driver source files,
but also the changes to all other kernel (header) files since the last revision.
The latter changes are sometimes even larger in size and effect than changes to the driver.
For example, the introduction of the kernel feature \verb|CONFIG_BRANCH_TRACER|
(profiling of unlikely and likely branches in the code by code instrumentation)
added several lines of auxiliary variables per \verb|if| statement,
and this additional code appears as new code in the next revision
that was made for each driver after the feature was introduced.
Our benchmark set of verification tasks has
an average of 688 changed lines of source code
between subsequent revisions.
Our results, presented in the following,
show that precision reuse
is quite insensitive to such large differences between revisions.

\subsection{Setup}
All experiments were performed on machines
with a 3.4\,GHz Quad Core CPU (Intel Core i7-2600) and 32\,GB of RAM.
We used Ubuntu~12.04 (64-bit) with Linux~3.2 and OpenJDK~1.7.
We used \cpachecker, revision $7537$.
The predicate analysis uses MathSAT~5.2.3 as SMT solver.
Each verification run was limited to 15 minutes of run-time and 15\,GB of RAM;
the Java heap size was limited to 10\,GB.
The run-time that we report refers to the total CPU time of the verification tool
(including startup and reading/writing of program-precision files),
and is given in seconds with two significant digits.
This is a similar environment to the community-agreed setting of
SV-COMP'13.
The size of code differences between two revisions of one program
is given as the number of differing lines excluding whitespace changes
(calculated with \verb+diff --ignore-all-space | diffstat+).

\subsection{Results}
We experiment with the reuse of precisions
across a sequence of different revisions of a program.
For this we start the verification of the first revision with the empty precision,
dump the generated precision
and use it as the initial precision for the verification of the second revision.
The final precision of the second verification run
is the input precision for the verification of the third revision and so on.
We compare the time needed for this process against the time that is needed
for verifying all the revisions individually
(using the empty precision as the input for each run
and without generating program-precision files).

\begin{table}
\caption{Results for verifying driver \texttt{dvb-usb-az6007} using predicate analysis}
\label{tbl:single-driver}
\centering
\vspace{1mm}
\scalebox{0.88}{
\begin{tabular}{lrr|grrp{5mm}|drrp{10mm}}
\toprule

& \multirow{2}{*}{\begin{sideways}\makebox[5.8em][l]{n-th Rev.}\end{sideways}}
& \multirow{2}{*}{\begin{sideways}\makebox[5.8em][l]{Diff. Lines}\end{sideways}}
& \multicolumn{4}{c|}{\emph{without} Precision Reuse} 
& \multicolumn{4}{c}{\emph{with} Precision Reuse} \\
Spec.\begin{sideways}\makebox[4.5em]{}\end{sideways}
&
&
& \multicolumn{1}{c}{\begin{rotate}{45}CPU Time \end{rotate}}
& \multicolumn{1}{c}{\begin{rotate}{45}Refinements \end{rotate}}
& \multicolumn{1}{c}{\begin{rotate}{45}Abstractions \end{rotate}}
& \multicolumn{1}{c|}{}
& \multicolumn{1}{c}{\begin{rotate}{45}CPU Time  \end{rotate}}
& \multicolumn{1}{c}{\begin{rotate}{45}Refinements \end{rotate}}
& \multicolumn{1}{c}{\begin{rotate}{45}Abstractions \end{rotate}}
& \multicolumn{1}{c}{}
\\
  \midrule

\multirow{5}{*}{08\_1a}      
    & 32 & -   & 9.1 & 2 & 1352 && 9.5 & 2 & 1352 \\ 
    & 33 & 593 & 9.5 & 2 & 1352 && 3.6 & 0 & 24 \\ 
    & 34 & 707 & 9.8 & 2 & 1352 && 3.6 & 0 & 24 \\ 
    & 35 & 478 & 9.3 & 2 & 1352 && 3.7 & 0 & 24 \\ 
    & 36 & 2   & 9.2 & 2 & 1352 && 3.9 & 0 & 24 \\ 
    \cmidrule(r){2-10}
    & & Total  & 47 & 10 & 6760 && 24 & 2 & 1448  \\
  \midrule

\multirow{5}{*}{32\_7a}      
   & 32 & -   & 6.5 & 27 & 186 && 6.6 & 27 & 186 \\ 
   & 33 & 752 & 7.2 & 28 & 210 && 4.5 & 1 & 48 \\ 
   & 34 & 961 & 8.0 & 29 & 234 && 4.8 & 1 & 48 \\ 
   & 35 & 462 & 7.9 & 29 & 234 && 4.4 & 0 & 24 \\ 
   & 36 & 2   & 7.9 & 29 & 234 && 4.4 & 0 & 24 \\ 
    \cmidrule(r){2-10}
    & & Total & 37 & 142 & 1098 && 25  & 29 & 330  \\
  \midrule

\multirow{5}{*}{39\_7a}      
   & 32 & -   & 58 & 10 & 8432 && 58 & 10 & 8432 \\ 
   & 33 & 595 & 58 & 10 & 8432 && 3.8 & 0 & 24 \\ 
   & 34 & 707 & 58 & 10 & 8432 && 4.1 & 0 & 24 \\ 
   & 35 & 462 & 60 & 10 & 8432 && 4.2 & 0 & 24 \\ 
   & 36 & 2   & 60 & 10 & 8432 && 4.2 & 0 & 24 \\ 
    \cmidrule(r){3-10}
    & & Total & 290 & 50 & 42160 && 75 & 10 & 8528 \\
\bottomrule

\end{tabular}
}
\end{table}

\subsubsection{Results for a Single Driver}
The results for a single driver (\verb|dvb-usb-az6007|) from the Linux kernel
are shown in Table~\ref{tbl:single-driver}.
There are five revisions for this driver,
and we show the verification of three specifications using predicate analysis.
The column ``Diff.~Lines'' shows the number of lines differing
in one revision compared to the previous revision.
The lines ``Total'' show the sum of the respective values
for all revisions with one specification.

As expected, the runtime for verifying the first revision
is not decreased by the reuse of precisions
(as there is no precision to reuse);
also, there is no significant overhead for writing the precision to the output file.
For the remaining revisions,
the runtime results show a clear improvement of performance
when reusing the precision from the previous revision.
This is achieved by almost completely eliminating the need for refinements,
and by lowering the number of (costly) boolean-abstraction computations considerably,
compared to the verification of the same program without precision reuse.
It is interesting to observe the second and third revisions
of this driver when verified against the specification \verb|32_7a|:
These two revisions affected the program source in a way
that made additional predicates necessary
(witnessed by the increase in the number of refinements from 27 to 29).
In such a case, the analysis with precision reuse
also has to perform refinements,
because these additional predicates are not yet known.
However, the 27~refinements that were necessary to discover predicates
without precision reuse,
are not necessary, because the results are read from the precision file.
Thus, the runtime is still much better than without precision reuse.

\begin{table*}
\caption{Results for predicate analysis (details for highlighted lines in Table~\ref{tbl:single-driver})}
\label{tbl:predicate}
\vspace{1mm}
\centering
\scalebox{0.88} {
\begin{tabular}{ll|rr|rr hih D{+}{+}{3.2}dr}
\toprule

  Device Driver
& Spec.
& \# Tasks
& Avg. Diff

& \multicolumn{2}{c}{Refinements}
& \multicolumn{3}{|c|}{CPU Time}
& \multicolumn{1}{c}{Solved}
& \multicolumn{1}{c}{Speedup}
& \multicolumn{1}{c}{Size of}

\\

&
&
& \multicolumn{1}{c|}{Lines}
& \multicolumn{1}{c}{w/o Reuse}
& \multicolumn{1}{c}{w/ Reuse}
& \multicolumn{1}{|c}{1st Rev.}
& \multicolumn{1}{c}{w/o Reuse}
& \multicolumn{1}{c|}{w/ Reuse}
& \multicolumn{1}{c}{Tasks}
& \multicolumn{1}{c}{}
& \multicolumn{1}{c}{Precision}

\\
\midrule

leds-bd2802 & 43\_1a & 4 & 426 & 210 & 6 & 220 & 1900 & 250 & 3+\mathbf{1} & 50 & 640 \\ 
  dp83640 & 39\_7a & 16 & 557 & 2256 & 140 & 590 & 11000 & 860 & 16 & 39 & 3516 \\ 
  mos7840 & 39\_7a & 57 & 621 & 27522 & 767 & 580 & 23000 & 1300 & 45+\mathbf{12} & 31 & 3307 \\ 
  dmx3191d & 39\_7a & 2 & 1597 & 104 & 57 & 640 & 1300 & 670 & 2 & 21 & 3321 \\ 
  dvb-usb-vp7045 & 39\_7a & 12 & 1001 & 356 & 44 & 41 & 1700 & 120 & 12 & 20 & 2680 \\ 
  leds-bd2802 & 08\_1a & 14 & 504 & 960 & 8 & 200 & 3300 & 360 & 14 & 20 & 471 \\ 
  ems\_usb & 39\_7a & 21 & 666 & 796 & 40 & 72 & 2400 & 190 & 21 & 20 & 2934 \\ 
\rowcolor{gray!20}
  dvb-usb-az6007 & 39\_7a & 5 & 353 & 50 & 10 & 58 & 290 & 75 & 5 & 15 & 1680 \\ 
  catc & 39\_7a & 22 & 893 & 1100 & 52 & 32 & 4100 & 340 & 22 & 13 & 3282 \\ 
  cp210x & 39\_7a & 71 & 256 & 24 & 12 & 770 & 1600 & 850 & 2+\mathbf{26} & 9.7 & 1539 \\ 
  spcp8x5 & 39\_7a & 37 & 481 & 4701 & 348 & 54 & 2400 & 310 & 37 & 9.4 & 1847 \\ 
  cxd2820r & 39\_7a & 23 & 468 & 624 & 42 & 3.1 & 4200 & 480 & 23 & 8.8 & 2380 \\ 
  i915 & 39\_7a & 79 & 842 & 3428 & 72 & 49 & 5800 & 770 & 78 & 8.0 & 3075 \\ 
  i2o\_scsi & 39\_7a & 6 & 454 & 381 & 64 & 29 & 230 & 54 & 6 & 7.6 & 2495 \\ 
  dmx3191d & 08\_1a & 2 & 1432 & 20 & 10 & 54 & 110 & 62 & 2 & 7.5 & 514 \\ 
  it87 & 39\_7a & 54 & 462 & 1358 & 37 & 20 & 5000 & 680 & 54 & 7.4 & 2091 \\ 
  dvb-usb-rtl28xxu & 39\_7a & 10 & 173 & 154 & 10 & 7.8 & 310 & 51 & 10 & 6.9 & 1820 \\ 
  sym53c500\_cs & 39\_7a & 19 & 468 & 1947 & 113 & 21 & 650 & 120 & 19 & 6.6 & 2634 \\ 
  arkfb & 39\_7a & 22 & 447 & 960 & 56 & 81 & 1200 & 260 & 20 & 6.4 & 2009 \\ 
  budget-patch & 39\_7a & 9 & 1669 & 205 & 27 & 12 & 200 & 44 & 9 & 6.0 & 2290 \\ 
  cp210x & 68\_1 & 14 & 538 & 954 & 162 & 330 & 4600 & 1100 & 14 & 5.4 & 938 \\ 
  mos7840 & 08\_1a & 60 & 795 & 722 & 15 & 46 & 3000 & 610 & 60 & 5.3 & 889 \\ 
  xilinx\_uartps & 39\_7a & 3 & 352 & 531 & 177 & 22 & 66 & 30 & 3 & 5.3 & 2248 \\ 
  farsync & 08\_1a & 5 & 984 & 159 & 30 & 18 & 100 & 33 & 5 & 5.0 & 815 \\ 
  it87 & 32\_7a & 59 & 463 & 860 & 25 & 17 & 2900 & 590 & 59 & 4.9 & 1696 \\ 
  ssu100 & 39\_7a & 28 & 337 & 791 & 44 & 35 & 830 & 200 & 28 & 4.7 & 2417 \\ 
  cp210x & 32\_1 & 14 & 219 & 1473 & 227 & 63 & 1100 & 310 & 14 & 4.3 & 693 \\ 
  mISDN\_core & 39\_7a & 59 & 1265 & 2651 & 50 & 20 & 2200 & 540 & 59 & 4.1 & 2691 \\ 
  it87 & 08\_1a & 59 & 478 & 603 & 14 & 18 & 2100 & 550 & 59 & 3.8 & 818 \\ 
  leds-bd2802 & 68\_1 & 4 & 463 & 57 & 16 & 38 & 170 & 75 & 4 & 3.4 & 1361 \\ 
  metro-usb & 39\_7a & 25 & 158 & 351 & 15 & 8.4 & 310 & 97 & 25 & 3.4 & 1417 \\ 
  i2c-algo-pca & 68\_1 & 7 & 477 & 238 & 35 & 8.6 & 68 & 26 & 7 & 3.3 & 917 \\ 
  vsxxxaa & 68\_1 & 2 & 1354 & 28 & 14 & 11 & 22 & 14 & 2 & 3.2 & 706 \\ 
  sil164 & 39\_7a & 3 & 383 & 54 & 18 & 12 & 37 & 20 & 3 & 3.2 & 1693 \\ 
  dp83640 & 08\_1a & 16 & 527 & 190 & 12 & 24 & 400 & 140 & 16 & 3.1 & 789 \\ 
  spcp8x5 & 68\_1 & 13 & 740 & 508 & 46 & 11 & 260 & 92 & 13 & 3.1 & 1385 \\ 
  leds-bd2802 & 32\_1 & 4 & 121 & 32 & 10 & 25 & 120 & 55 & 4 & 3.1 & 1135 \\ 
  cp210x & 08\_1a & 71 & 304 & 186 & 9 & 170 & 8200 & 2800 & 71 & 3.1 & 387 \\ 
  i915 & 08\_1a & 79 & 731 & 1264 & 20 & 69 & 1900 & 680 & 79 & 3.0 & 527 \\ 
  uartlite & 39\_7a & 9 & 326 & 198 & 22 & 11 & 98 & 40 & 9 & 3.0 & 2151 \\ 
  it87 & 43\_1a & 15 & 612 & 105 & 7 & 9.0 & 150 & 56 & 15 & 2.9 & 405 \\ 
  i2c-algo-pca & 32\_1 & 7 & 223 & 131 & 19 & 6.7 & 50 & 21 & 7 & 2.9 & 668 \\ 
  i915 & 32\_7a & 79 & 777 & 1184 & 24 & 5.9 & 1800 & 640 & 79 & 2.9 & 1020 \\ 
  mos7840 & 32\_7a & 60 & 615 & 779 & 49 & 6.4 & 2500 & 900 & 60 & 2.8 & 1902 \\ 
  cp210x & 32\_7a & 71 & 257 & 600 & 20 & 44 & 5500 & 2100 & 56+\mathbf{10} & 2.6 & 1248 \\ 
  lms283gf05 & 39\_7a & 13 & 458 & 320 & 23 & 11 & 140 & 62 & 13 & 2.6 & 1658 \\ 
  arkfb & 68\_1 & 6 & 706 & 136 & 48 & 95 & 300 & 170 & 6 & 2.6 & 2575 \\ 
\rowcolor{gray!20}
  dvb-usb-az6007 & 08\_1a & 5 & 356 & 10 & 2 & 9.5 & 47 & 24 & 5 & 2.6 & 312 \\ 
  twidjoy & 39\_7a & 2 & 1458 & 46 & 26 & 10 & 19 & 14 & 2 & 2.5 & 2159 \\ 
  wm831x-dcdc & 39\_7a & 34 & 286 & 133 & 4 & 4.8 & 470 & 190 & 34 & 2.5 & 1402 \\ 

\midrule
\multicolumn{12}{c}{\vspace{2mm}
\begin{minipage}[c][5\baselineskip][c]{1mm}\vdots\end{minipage}
\hfill
\begin{minipage}[b][5\baselineskip][c]{13cm}\centering For full results c.f. \url{http://www.sosy-lab.org/~dbeyer/cpa-reuse/predicate.html} \end{minipage}
\hfill
\begin{minipage}[c][5\baselineskip][c]{1mm}\vdots\end{minipage}} \\
\midrule

tcm\_loop & 32\_7a & 41 & 259 & 58 & 3 & 4.0 & 160 & 160 & 41 & 1.0 & 614 \\ 
  mt2266 & 32\_7a & 5 & 748 & 3 & 1 & 2.3 & 13 & 12 & 5 & 1.0 & 565 \\ 
  adl\_pci7432 & 39\_7a & 13 & 122 & 46 & 4 & 2.5 & 32 & 32 & 13 & 1.0 & 816 \\ 
  slram & 08\_1a & 9 & 563 & 60 & 6 & 3.5 & 34 & 33 & 9 & 1.0 & 490 \\ 
  spi\_ks8995 & 32\_7a & 4 & 516 & 12 & 4 & 2.8 & 11 & 11 & 4 & 1.0 & 828 \\ 
  drbd & 08\_1a & 96 & 2657 & 96 & 1 & 9.0 & 870 & 860 & 96 & 1.0 & 245 \\ 
  mtdoops & 08\_1a & 41 & 264 & 47 & 4 & 4.4 & 110 & 110 & 41 & 1.0 & 539 \\ 
  farsync & 32\_7a & 9 & 889 & 0 & 0 & 4.2 & 49 & 49 & 9 & 1.0 &   2 \\ 
  rtc-max6902 & 32\_1 & 5 & 564 & 5 & 1 & 2.9 & 14 & 14 & 5 & 1.0 & 221 \\ 
  adl\_pci7432 & 08\_1a & 13 & 122 & 23 & 2 & 2.4 & 31 & 31 & 13 & 1.0 & 604 \\ 
  wl12xx\_sdio & 32\_7a & 38 & 261 & 42 & 3 & 3.2 & 130 & 130 & 38 & 1.0 & 776 \\ 
  ar7part & 43\_1a & 2 & 220 & 3 & 3 & 2.0 & 4.2 & 4.2 & 2 & 1.0 & 277 \\ 
  i915 & 43\_1a & 79 & 746 & 0 & 0 & 6.2 & 640 & 640 & 79 & .99 &   2 \\ 
  mISDN\_core & 43\_1a & 26 & 2079 & 156 & 6 & 7.0 & 190 & 190 & 26 & .99 & 223 \\ 
  mt2266 & 08\_1a & 5 & 725 & 5 & 1 & 2.6 & 13 & 13 & 5 & .99 & 557 \\ 
  rtc-max6902 & 43\_1a & 5 & 562 & 4 & 1 & 2.7 & 13 & 13 & 5 & .99 & 219 \\ 
  rtc-pcf2123 & 32\_7a & 9 & 747 & 27 & 9 & 2.9 & 28 & 28 & 9 & .98 & 1252 \\ 
  i2c-matroxfb & 43\_1a & 5 & 409 & 8 & 2 & 2.5 & 12 & 13 & 5 & .98 & 257 \\ 
  keyspan\_remote & 32\_1 & 3 & 285 & 3 & 1 & 2.7 & 7.7 & 8.0 & 3 & .98 & 297 \\ 
  wl12xx\_sdio & 08\_1a & 38 & 258 & 38 & 2 & 3.2 & 120 & 130 & 38 & .97 & 579 \\ 
  i2c-matroxfb & 08\_1a & 7 & 565 & 7 & 2 & 2.6 & 18 & 19 & 7 & .96 & 283 \\ 
  ads7871 & 08\_1a & 10 & 265 & 10 & 1 & 2.3 & 22 & 23 & 10 & .96 & 245 \\ 
  magellan & 32\_7a & 2 & 1267 & 10 & 9 & 3.9 & 7.4 & 7.6 & 2 & .96 & 1209 \\ 
  slram & 32\_7a & 9 & 625 & 34 & 16 & 2.7 & 30 & 36 & 9 & .83 & 1618 \\ 
  mos7840 & 43\_1a & 25 & 1018 & 518 & 1 & 390 & 980 & 4800 & 11+\mathbf{7} & .21 & 3965 \\ 

\midrule

  \textbf{Sum}
&
& 4193 
&  
& 80280 
& 5034 
& 5800 
& 130000 
& 40000 
& 4001+\mathbf{56} 
&  
& 242197 
\\

  \textbf{Average}
&
& 16 
& 688 
& 321 
& 20 
& 23 
& 520 
& 160 
& 16 
& 3.7 
& 969 
\\

\bottomrule
\end{tabular}
}
\end{table*}

\begin{table*}
\caption{Results for explicit-value analysis}
\label{tbl:explicit}
\vspace{1mm}
\centering
\scalebox{0.88} {
\begin{tabular}[m]{ll|rr|rr hih D{+}{+}{3.2}dr}
\toprule

  Device Driver
& Spec.
& \# Tasks
& Avg. Diff.

& \multicolumn{2}{c}{Refinements}
& \multicolumn{3}{|c|}{CPU Time}
& \multicolumn{1}{c}{Solved}
& \multicolumn{1}{c}{Speedup}
& \multicolumn{1}{c}{Size of}

\\

&
&
& \multicolumn{1}{c|}{Lines}
& \multicolumn{1}{c}{w/o Reuse}
& \multicolumn{1}{c}{w/ Reuse}
& \multicolumn{1}{|c}{1st Rev.}
& \multicolumn{1}{c}{w/o Reuse}
& \multicolumn{1}{c|}{w/ Reuse}
& \multicolumn{1}{c}{Tasks}
& \multicolumn{1}{c}{}
& \multicolumn{1}{c}{Precision}

\\
\midrule

cfag12864b & 08\_1a & 4 & 326 & 36 & 9 & 75 & 290 & 89 & 4 & 15 & 4846 \\ 
  cfag12864b & 32\_1 & 2 & 48 & 14 & 7 & 71 & 140 & 76 & 2 & 13 & 3175 \\ 
  cfag12864b & 39\_7a & 4 & 414 & 49 & 13 & 240 & 1100 & 310 & 4 & 12 & 20606 \\ 
  mISDN\_core & 39\_7a & 59 & 1265 & 738 & 15 & 25 & 1900 & 490 & 59 & 4.0 & 34859 \\ 
  cfag12864b & 32\_7a & 4 & 369 & 37 & 10 & 71 & 350 & 140 & 4 & 3.8 & 10573 \\ 
  it87 & 39\_7a & 54 & 462 & 478 & 10 & 5.7 & 540 & 250 & 54 & 2.2 & 3324 \\ 
  mISDN\_core & 68\_1 & 26 & 2481 & 52 & 2 & 6.8 & 410 & 200 & 26 & 2.1 & 958 \\ 
  tcm\_loop & 39\_7a & 41 & 263 & 517 & 14 & 8.1 & 360 & 180 & 41 & 2.0 & 15686 \\ 
  budget-patch & 43\_1a & 5 & 1239 & 20 & 4 & 6.7 & 37 & 22 & 5 & 2.0 & 4135 \\ 
  mISDN\_core & 32\_7a & 59 & 1179 & 202 & 7 & 10 & 860 & 440 & 59 & 2.0 & 26517 \\ 
  sil164 & 39\_7a & 3 & 383 & 18 & 6 & 6.5 & 18 & 13 & 3 & 1.9 & 9100 \\ 
  com20020\_cs & 39\_7a & 2 & 524 & 18 & 9 & 6.2 & 12 & 9.4 & 2 & 1.9 & 11896 \\ 
  uartlite & 39\_7a & 9 & 326 & 63 & 7 & 6.4 & 52 & 31 & 9 & 1.9 & 15106 \\ 
  i2o\_scsi & 39\_7a & 6 & 454 & 55 & 10 & 6.4 & 38 & 23 & 6 & 1.8 & 10457 \\ 
  wl12xx\_sdio & 39\_7a & 38 & 266 & 372 & 11 & 6.6 & 240 & 130 & 38 & 1.8 & 2869 \\ 
  ems\_usb & 39\_7a & 21 & 666 & 199 & 10 & 7.0 & 140 & 82 & 21 & 1.8 & 6129 \\ 
  slram & 68\_1 & 5 & 511 & 25 & 5 & 5.7 & 28 & 18 & 5 & 1.8 & 7030 \\ 
  mISDN\_core & 08\_1a & 59 & 1532 & 118 & 2 & 11 & 730 & 420 & 59 & 1.8 & 1972 \\ 
  slram & 39\_7a & 9 & 599 & 72 & 9 & 5.5 & 51 & 31 & 9 & 1.8 & 7567 \\ 
  cx231xx-dvb & 39\_7a & 13 & 577 & 127 & 10 & 6.2 & 87 & 53 & 13 & 1.7 & 6854 \\ 
  it87 & 32\_7a & 59 & 463 & 299 & 11 & 4.9 & 480 & 270 & 59 & 1.7 & 15657 \\ 
  dvb-usb-az6007 & 39\_7a & 5 & 353 & 45 & 9 & 7.0 & 35 & 23 & 5 & 1.7 & 14225 \\ 
  dvb-usb-rtl28xxu & 39\_7a & 10 & 173 & 90 & 9 & 6.3 & 66 & 41 & 10 & 1.7 & 8771 \\ 
  it87 & 08\_1a & 59 & 478 & 229 & 3 & 4.7 & 430 & 250 & 59 & 1.7 & 535 \\ 
  arkfb & 39\_7a & 22 & 447 & 132 & 7 & 7.1 & 180 & 110 & 22 & 1.7 & 4375 \\ 
  dp83640 & 39\_7a & 16 & 557 & 176 & 11 & 6.9 & 110 & 69 & 16 & 1.7 & 9995 \\ 
  dvb-usb-vp7045 & 39\_7a & 12 & 1001 & 110 & 11 & 6.5 & 78 & 48 & 12 & 1.7 & 13195 \\ 
  tdo24m & 39\_7a & 12 & 536 & 74 & 7 & 6.3 & 68 & 43 & 12 & 1.7 & 5015 \\ 
  i2c-matroxfb & 39\_7a & 7 & 617 & 51 & 8 & 5.3 & 35 & 24 & 7 & 1.7 & 5978 \\ 
  catc & 39\_7a & 22 & 893 & 246 & 13 & 6.3 & 150 & 93 & 22 & 1.7 & 6604 \\ 
  cp210x & 32\_1 & 14 & 219 & 56 & 4 & 6.8 & 94 & 60 & 14 & 1.7 & 4175 \\ 
  cp210x & 39\_7a & 71 & 256 & 456 & 8 & 6.2 & 460 & 280 & 71 & 1.7 & 1363 \\ 
  budget-patch & 39\_7a & 9 & 1669 & 98 & 13 & 6.0 & 57 & 37 & 9 & 1.7 & 9617 \\ 
  xilinx\_uartps & 39\_7a & 3 & 352 & 21 & 7 & 5.6 & 16 & 12 & 3 & 1.6 & 11926 \\ 
  panasonic-laptop & 39\_7a & 16 & 410 & 104 & 7 & 4.9 & 78 & 50 & 16 & 1.6 & 2549 \\ 
  slram & 32\_1 & 5 & 450 & 20 & 4 & 5.0 & 24 & 17 & 5 & 1.6 & 3330 \\ 
  sil164 & 32\_7a & 3 & 486 & 15 & 6 & 5.3 & 16 & 12 & 3 & 1.6 & 8326 \\ 
  sym53c500\_cs & 39\_7a & 19 & 468 & 175 & 10 & 6.4 & 120 & 75 & 19 & 1.6 & 5914 \\ 
  spcp8x5 & 39\_7a & 37 & 481 & 273 & 9 & 6.2 & 230 & 140 & 37 & 1.6 & 1962 \\ 
  wm831x-dcdc & 39\_7a & 34 & 286 & 133 & 4 & 4.5 & 180 & 110 & 34 & 1.6 & 593 \\ 
  dmx3191d & 39\_7a & 2 & 1597 & 14 & 8 & 6.2 & 12 & 10 & 2 & 1.6 & 9896 \\ 
  ssu100 & 39\_7a & 28 & 337 & 209 & 9 & 6.7 & 170 & 110 & 28 & 1.6 & 5041 \\ 
  dvb-usb-vp7045 & 32\_1 & 2 & 1806 & 12 & 6 & 5.9 & 12 & 10 & 2 & 1.6 & 3961 \\ 
  wm831x-dcdc & 68\_1 & 3 & 128 & 10 & 4 & 6.1 & 17 & 13 & 3 & 1.5 & 8991 \\ 
  metro-usb & 39\_7a & 25 & 158 & 175 & 7 & 5.5 & 120 & 83 & 25 & 1.5 & 4102 \\ 
  abyss & 39\_7a & 3 & 2202 & 26 & 10 & 6.4 & 19 & 15 & 3 & 1.5 & 13496 \\ 
  pcc-cpufreq & 39\_7a & 3 & 554 & 21 & 7 & 4.4 & 13 & 10 & 3 & 1.5 & 6689 \\ 
  mos7840 & 39\_7a & 57 & 621 & 416 & 9 & 6.8 & 410 & 280 & 57 & 1.5 & 1290 \\ 
  tdo24m & 32\_7a & 12 & 586 & 60 & 8 & 5.1 & 64 & 45 & 12 & 1.5 & 4445 \\ 
  keyspan\_remote & 39\_7a & 7 & 929 & 43 & 7 & 5.0 & 31 & 23 & 7 & 1.5 & 7903 \\ 

\midrule
\multicolumn{12}{c}{\vspace{2mm}
\begin{minipage}[c][5\baselineskip][c]{1mm}\vdots\end{minipage}
\hfill
\begin{minipage}[b][5\baselineskip][c]{13cm}\centering For full results c.f. \url{http://www.sosy-lab.org/~dbeyer/cpa-reuse/explicit.html} \end{minipage}
\hfill
\begin{minipage}[c][5\baselineskip][c]{1mm}\vdots\end{minipage}} \\
\midrule

videobuf-vmalloc & 08\_1a & 31 & 363 & 31 & 1 & 2.6 & 79 & 79 & 31 & 1.0 &  57 \\ 
  intel\_vr\_nor & 08\_1a & 10 & 275 & 10 & 1 & 2.6 & 24 & 24 & 10 & .99 & 1130 \\ 
  uio\_sercos3 & 32\_7a & 5 & 886 & 7 & 3 & 2.4 & 13 & 13 & 5 & .99 & 2842 \\ 
  abyss & 43\_1a & 3 & 1465 & 0 & 0 & 2.8 & 9.1 & 9.0 & 3 & .99 &   0 \\ 
  rtc-pcf2123 & 43\_1a & 2 & 59 & 2 & 1 & 3.0 & 5.8 & 5.8 & 2 & .99 & 1351 \\ 
  vsxxxaa & 43\_1a & 2 & 786 & 2 & 1 & 2.9 & 5.3 & 5.5 & 2 & .99 & 699 \\ 
  twidjoy & 08\_1a & 2 & 1222 & 2 & 1 & 2.7 & 4.9 & 5.0 & 2 & .99 & 1154 \\ 
  spcp8x5 & 43\_1a & 13 & 897 & 13 & 1 & 3.4 & 48 & 48 & 13 & .99 & 652 \\ 
  i2c-algo-pca & 08\_1a & 14 & 480 & 14 & 1 & 2.4 & 35 & 35 & 14 & .99 & 173 \\ 
  i915 & 08\_1a & 79 & 731 & 93 & 1 & 4.5 & 510 & 510 & 79 & .99 & 463 \\ 
  farsync & 32\_7a & 9 & 889 & 0 & 0 & 3.1 & 29 & 30 & 9 & .99 &   0 \\ 
  magellan & 32\_7a & 2 & 1267 & 2 & 2 & 2.6 & 5.0 & 5.1 & 2 & .98 & 1557 \\ 
  comedi\_bond & 08\_1a & 13 & 98 & 13 & 1 & 2.4 & 29 & 30 & 13 & .98 & 173 \\ 
  mISDN\_core & 32\_1 & 26 & 388 & 26 & 1 & 6.4 & 180 & 180 & 26 & .98 & 308 \\ 
  mtdoops & 43\_1a & 20 & 323 & 20 & 1 & 2.5 & 52 & 53 & 20 & .98 & 223 \\ 
  rtc-max6902 & 32\_7a & 9 & 829 & 7 & 3 & 2.5 & 23 & 23 & 9 & .98 & 2046 \\ 
  cxd2820r & 32\_7a & 23 & 492 & 32 & 3 & 3.4 & 96 & 98 & 23 & .98 & 1094 \\ 
  cfag12864b & 43\_1a & 2 & 74 & 2 & 1 & 2.3 & 4.5 & 4.5 & 2 & .97 & 399 \\ 
  twidjoy & 32\_7a & 2 & 1268 & 2 & 2 & 2.6 & 4.9 & 5.1 & 2 & .97 & 1557 \\ 
  abyss & 32\_7a & 4 & 2025 & 2 & 2 & 2.8 & 13 & 14 & 4 & .96 & 2774 \\ 
  cxd2820r & 08\_1a & 23 & 451 & 23 & 1 & 3.1 & 92 & 97 & 23 & .95 & 119 \\ 
  spaceorb & 32\_7a & 2 & 1226 & 2 & 2 & 2.8 & 5.0 & 5.3 & 2 & .94 & 1557 \\ 
  mISDN\_core & 43\_1a & 26 & 2079 & 26 & 1 & 5.8 & 170 & 180 & 26 & .92 & 311 \\ 
  dmx3191d & 32\_7a & 2 & 1608 & 3 & 3 & 3.0 & 7.3 & 7.8 & 2 & .89 & 3246 \\ 
  drbd & 08\_1a & 96 & 2657 & 96 & 1 & 10 & 950 & 1100 & 96 & .89 &  89 \\ 

\midrule

  \textbf{Sum}
&
& 4193 
&  
& 13313 
& 911 
& 1500 
& 27000 
& 20000 
& 4191 
&  
& 756465 
\\

  \textbf{Average}
&
& 16 
& 688 
& 52 
& 4 
& 5.8 
& 100 
& 76 
& 16 
& 1.4 
& 2932 
\\

\bottomrule
\end{tabular}
}
\end{table*}

\subsubsection{Results for all Device Drivers and Specifications}
Tables~\ref{tbl:predicate} and~\ref{tbl:explicit}
show the results of verifying all revisions of all \numofdrivers~device drivers
against all appropriate specifications,
with predicate analysis and explicit analysis, respectively.
Due to space reasons we restrict these tables
to the 50~best and 25~worst cases
out of the total \numofdriverspecs~driver/specification pairs
(sorted by column ``Speedup'').
The complete tables are available on the supplementary webpage.

The columns ``CPU Time''
show the total time used by the model checker
to verify all revisions of the device driver against the given specification
(excluding the revisions for which it failed due to a timeout
or an out-of-memory condition).
The column ``CPU Time 1st Rev.'' shows the time
needed for verifying the first revision
(this is the same with and without reuse).
The column ``Solved Tasks'' shows the number of successfully verified revisions
out of the total number of revisions for this driver
(the remaining cases were either timeout or out-of-memory,
there were no incorrect verification results).
If the value in this column is of format ``$N+M$'',
this means that without precision reuse only $N$ revisions could be verified,
whereas with precision reuse $N+M$ revisions could be verified;
otherwise the number of successfully verified revisions is the same.
There were no cases where a revision could be verified without reuse,
but not with precision reuse.
The column ``Speedup'' gives the average speedup
for the task of verifying a single revision of the driver
when a precision from a previous revision is available for reuse,
as opposed to the case where no information is available for reuse.
The verification time of the first revision of each driver
is not taken into account for calculating the speedup,
in order to make this value independent from the number of revisions per driver
(otherwise a driver with more revisions
would in general show a higher speedup
because the cost of the verification of the first revision
is less relevant).
We also excluded from calculating the speedup such revisions
that could not be verified by one or both of the two configurations
(without and with reuse).
In the last column, we report the size in bytes of the final program-precision file
that was produced during the verification of the revisions of this driver.
Note that our file format is purely text-based,
thus, this number gives a coarse over-approximation of the amount of information
that is reused between verification runs.
The two highlighted rows show the driver \verb|dvb-usb-az6007|,
for which further details are available in Table~\ref{tbl:single-driver}
(the two lines here correspond to the lines labeled ``Total''
in the previous table).
The bottom rows of the table report the sum
and the average of the respective values per driver/specification pair.

Precision reuse not only increases the efficiency, but also the effectiveness:
For five~pairs of driver and specification,
the number of successfully solved verification tasks was increased
by our approach (for predicate analysis).
This may happen if an early revision of a driver is verifiable,
and a later revision would need more than $900$\,s to be verified.
With precision reuse, the verification of the later revision is easier,
because a large part of the precision is given as input;
often up to the point that it actually can be verified successfully.
The maximum speedup for predicate analysis is~50,
and for~77 out of \numofdriverspecs driver/specification pairs
the speedup is at least~two.

We also list all negative results: there are only a few.
The last lines of the tables report the few cases for which the verification
with precision reuse takes a bit more time than without.
Most of these cases have a rather low average CPU time per revision,
and in almost all of these cases, the performance drop is not worse than~$5$\,\%.
There is one case for which the time for verification with precision reuse
is significantly higher
(the last line in Table~\ref{tbl:predicate}, driver \verb|mos7840| with specification \verb|43_1a|).
However, note that precision reuse increased the number of successfully solved tasks
from~11~to~18 for this case.
We generally consider an increase in the number of solved programs
to be more important than a performance difference.
The verification of the same driver against the other specifications actually shows nice speedups
(e.g., third line; also with increase of solved tasks).

The total time that the predicate analysis used for successfully verifying
4001~verification tasks without precision reuse was $130\,000$\,s,
whereas with precision reuse a total of 4057~verification tasks (56 more) were verified
in only $40\,000$\,s, less than a third of the time.
This gives evidence of the significant performance improvement of our approach.

\subsubsection{Size of Precision}
The size of the precision that is necessary to be stored between subsequent verification runs
is small: usually just a few KBs
in our uncompressed plain-text format.
The average size for predicate analysis is~1\,KB (max: 4\,KB);
for explicit analysis it is~3\,KB (max: 35\,KB).

The total amount of precision storage that was necessary for verifying all \numoftasks~verification tasks
was 236\,KB for predicate analysis and 738\,KB for explicit analysis,
which is orders of magnitude less compared to the size of the source-code.

\subsubsection{Scaling with Larger Changes}
\begin{table}
\caption{Results for considering all revisions versus considering only every 4th revision}
\label{tbl:4th-revs}
\vspace{1mm}
\centering
\scalebox{0.88}{
\hspace{-2mm}
\begin{tabular}{lr|rr|rr cD{+}{+}{4.2}}
\toprule
  Analysis
& Revs.
& \# Tasks
& Avg.
& \multicolumn{2}{c|}{CPU Time}
& Speedup\!\!\!
& \multicolumn{1}{c}{Solved}
\\

&
&
& \multicolumn{1}{c|}{Diff.}
& \multicolumn{1}{c}{w/o}
& \multicolumn{1}{c|}{w/}
&
\\

&
&
& \multicolumn{1}{c|}{Lines}
& \multicolumn{1}{c}{Reuse}
& \multicolumn{1}{c|}{Reuse}
&
\\
\midrule
\multirow{2}{*}{Predicate\!\!}
    & All &  4193 & 688      & 130000            & 40000             & 3.7         & 4001+56 \\
    & 4th &  1090 & 1579     & 34000             & 14000             & 3.2         & 1045+12 \\
\midrule
\multirow{2}{*}{Explicit}
    & All & 4193  & 688       & 27000             & 20000             & 1.4         & 4191 \\
    & 4th & 1090  & 1579      & 6300              & 5100              & 1.3         & 1090 \\
\bottomrule
\end{tabular}
}
\end{table}
As explained above,
the changes between subsequent revisions in our benchmark set
are already rather large
(affecting 688~lines on average) compared to typical developer commits.
To find out how our approach scales with the size of changes per revision (change-size sensitivity),
we created verification problems with even more changes:
we consider only every 4th revision per driver/specification pair 
as an alternative benchmark set.
Thus, the difference between two revisions in this benchmark set
combines the differences of four actual driver revisions.

Table~\ref{tbl:4th-revs} shows the result for this experiment
in the lines that are marked ``4th'' in column ``Revs.''
(the lines marked ``All'' show the previous results for comparison).
The average size of differences between revisions
increased from 688 to 1579 lines.
As expected, the speedup decreased,
but only from $3.7$ to $3.2$ for predicate analysis,
and from $1.4$ to $1.3$ for explicit analysis.
This shows that our approach copes well even with massive changes to the analyzed code.

\subsection{Threats to Validity}
To have a significant experimental basis,
we created a huge set of \numoftasks benchmark verification tasks.
To derive highly credible test data,
and instead of relying on random or artificial benchmarks, 
our selection of verification tasks
is based on hundreds of actual source code commits
to \numofdrivers different Linux device drivers.
The characteristics of systems software, in particular kernel device drivers,
might be similar and could have an impact on validity,
but,
Linux driver verification is important
enough to be representative on its own~\cite{LDV12}.
After all, there is the Linux Driver Verification Program
of the Linux Verification Center~\cite{LDV}
and also Microsoft dedicates considerable amounts of resources
to Windows driver verification~\cite{SLAM}.
We used an experimental setup and environment
that is virtually identical to the infrastructure
for the competition on software verification (community-agreed).
Precision reuse has a different impact on different abstract domains.
We included two totally different analysis approaches in our experimental evaluation:
a symbolic and an explicit model-checking approach.
Our experiments are not operating on one particular specification
to check, but we rather consider six different, real-world specifications, 
with all showing a considerable speedup.

\section{Conclusion}

We propose to use abstraction precisions as reusable verification facts.
Precisions are easy to extract from model checkers that automatically
construct an abstract model of the program (abstract interpretation).
Precisions are tool-independent and it is easy for successive verification runs
to read and use precisions;
their memory footprint is small.

We present an extensive collection of verification tasks for 
benchmarking approaches for regression verification
that is derived from industrial code, namely, the Linux kernel.
Our benchmark consists of \numoftasks single verification problems and
is publicly available on our supplementary web page.

Our experiments confirm that the reuse of precisions has a significant effect
on the verification process.
The approach drastically improves the performance on most verification problems,
and if not successful, it does not have a noticeable negative impact.
Besides improving the performance,
we sometimes even solve verification problems that
were not solvable before in the given time and memory resources.

The technical insight of our approach is that reusing precisions
drastically reduces the number of CEGAR iterations (refinements),
and therefore the effort spent on analyzing spurious counterexamples
and reconstructing abstract states for refined parts of the system.
Precisions are precious intermediate results 
that are difficult to discover, and which define the abstraction level of the abstract model.
Thus, the work on discovering the abstract model is significantly reduced in later verification runs.
Because the information that we reuse does not depend on source-code details,
our approach is less sensitive to changes in the source code, compared to other approaches.
Precision reuse is applicable to all verification approaches that are based on abstraction
and automatically compute the precision of the abstract model
(this includes \cegar-based approaches and abstract interpretations).

As a result of the experiments for this paper,
an unknown bug in the Linux kernel was found
and a fix was submitted to the maintainers by the LDV team.

\bibliography{dbeyer,sw,tah}
\bibliographystyle{abbrv}

\end{document}